\documentclass[prl,twocolumn]{revtex4}
\usepackage{graphicx}
\usepackage{amssymb}
\usepackage{amsmath}

\begin{document}

\title{Remote Entanglement between a Single Atom and a Bose-Einstein Condensate}
\author{M. Lettner}
\author{M. M\"{u}cke}
\author{S. Riedl}
\author{C. Vo}
\author{C. Hahn}
\author{S. Baur}
\author{J. Bochmann}
\author{S. Ritter}
\author{S. D\"{u}rr}
\author{G. Rempe}
\affiliation{Max-Planck-Institut f{\"u}r Quantenoptik, Hans-Kopfermann-Stra{\ss}e 1, 85748 Garching, Germany}

\begin{abstract}
Entanglement between stationary systems at remote locations is a key resource for quantum networks. We report on the experimental generation of remote entanglement between a single atom inside an optical cavity and a Bose-Einstein condensate (BEC). To produce this, a single photon is created in the atom-cavity system, thereby generating atom-photon entanglement. The photon is transported to the BEC and converted into a collective excitation in the BEC, thus establishing matter-matter entanglement. After a variable delay, this entanglement is converted into photon-photon entanglement. The matter-matter entanglement lifetime of 100~$\mu$s exceeds the photon duration by two orders of magnitude. The total fidelity of all concatenated operations is 95~\%. This hybrid system opens up promising perspectives in the field of quantum information.
\end{abstract}

\maketitle

While quantum-mechanical entanglement is always created locally, remote entanglement between stationary qubits is an essential ingredient for quantum networks \cite{kimble:08}. This can be achieved by transferring photons between different nodes of the network. The functionality of such a network relies crucially on the faithful conversion of quantum states between flying and stationary qubits. Previous experiments on remote matter-matter entanglement employed either inherently probabilistic schemes for qubit entanglement \cite{matsukevich:06, moehring:07, chou:07, yuan:08} or spin squeezing instead of qubits \cite{julsgaard:01}. Our work follows an alternative approach as it implements a protocol in which all steps can be made reversible and deterministic. A single atom inside an optical cavity is used for the controlled generation of a single photon and of atom-photon entanglement \cite{wilk:07, weber:09}. After creating the single photon, it is transported through an optical fiber to the BEC and stored there, thus yielding remote matter-matter entanglement. Raman storage is achieved in a scheme based on electromagnetically induced transparency (EIT) \cite{fleischhauer:05, chaneliere:05, eisaman:05}. The BEC is well suited for this purpose because the absence of thermal motion allows for long storage times, the large optical depth allows for high write-read efficiencies, and excellent internal-state preparation allows for high-fidelity storage of a qubit in atomic spin states. The matter-matter entanglement is maintained for an adjustable time and then converted into photon-photon entanglement by independently creating two single photons, one from the atom and one from the BEC. The observed fidelity of 95 \% with the expected Bell state demonstrates the excellent overall performance of this protocol in which entanglement created in the source is nearly perfectly sustained throughout all entangling and disentangling operations. The matter-matter entanglement survives for 100 $\mu$s which is two orders of magnitude longer than the intrinsic times scales for photon generation and photon transport. The atom-cavity system coupled to the many-body quantum gas forms a novel hybrid system with promising applications in quantum communication, quantum metrology, and quantum computation \cite{kimble:08, briegel:98, duan:04, giovannetti:04, rispe:1010.0037}.

\begin{figure*}[t]
\includegraphics[width=0.98\textwidth]{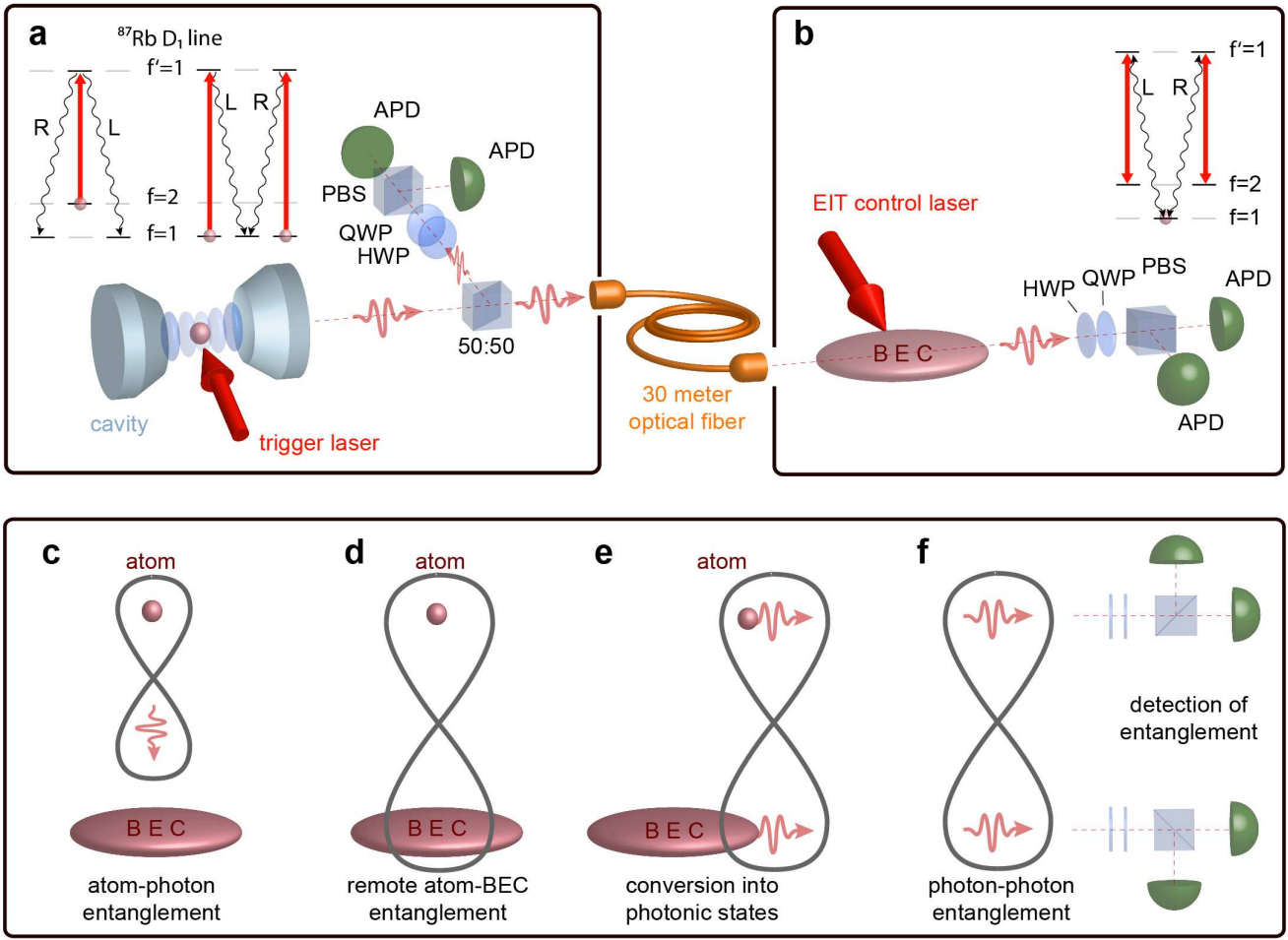}
\caption{
\label{fig-scheme}
{\bf Scheme of the experiment.} {\bf a} A single atom inside an optical cavity serves as a source for polarization-entangled pairs of single photons. After impinging on a 50:50 beam splitter, the photon polarization can either be measured immediately or the photon can be transported in an optical fiber to a different laboratory, {\bf b}, where a BEC serves as an EIT-type quantum memory. After storage, the photon is retrieved and its polarization is measured. Some optical components, such as mode-filtering fiber, filter cavity, etc., are not shown here. {\bf c}~The experimental sequence begins with a trigger pulse which illuminates the single atom to generate a single photon. This process creates a maximally-entangled Bell state of the photon polarization and the atomic spin state, see Eq.\ (\ref{at-ph}). {\bf d}~After transport in an optical fiber, the single photon is stored as a spin-wave excitation in the BEC. This establishes remote atom-BEC entanglement, see Eq.\ (\ref{at-BEC}). {\bf e}~After arbitrary delays, which can be chosen independently for each atomic system, the matter-matter entanglement is converted into photon-photon entanglement, see Eq.\ (\ref{ph-ph}). {\bf f}~Finally, the polarizations of both photons are measured.
}
\end{figure*}

The experimental setup and sequence are sketched in Fig.\ \ref{fig-scheme}. A light pulse from a laser triggers the emission of a single-photon pulse from a single $^{87}$Rb atom in a high-finesse optical cavity. This creates the entangled atom-photon state
\begin{equation}
\label{at-ph}
|\psi_{\rm at\otimes ph}\rangle = (|1,1\rangle\otimes|L\rangle-|1,-1\rangle\otimes|R\rangle)/\sqrt2
.\end{equation}
$|L\rangle$ and $|R\rangle$ refer to left- and right-hand circular photon polarization, respectively. The atomic spin state is $|f,m_f\rangle$ with hyperfine quantum numbers $f,m_f$. The shape of the single-photon pulse is controlled by the shape of the trigger laser pulse. It is chosen such that the total duration of a single-photon pulse is $0.45$ $\mu$s (full width at half maximum: $0.18$ $\mu$s). The emitted photons exhibit sub-Poissonian statistics with $g^{(2)}(0)=(1.0\pm 0.4)\% \ll 1$.

The photon is transported in a 30 meter long optical fiber to a different laboratory, where it is absorbed in an $^{87}$Rb BEC using Raman transfer. This EIT-type scheme converts the single photon into a single magnon, which is a quasi-particle of a collective spin-wave excitation in the BEC. The single magnon is described by the state vector $|f,m_f\rangle=(1/\sqrt N) \sum_{j=1}^N |f,m_f\rangle_j \otimes |\chi_2\rangle_j \bigotimes_{k=1,k\neq j}^N |1,0\rangle_k \otimes |\chi_1\rangle_k$ where $|...\rangle_j$ describes the state of the $j$-th atom in the BEC and $\chi_1$ and $\chi_2$ are spatial wavefunctions. $f,m_f$ are the hyperfine quantum numbers of the atom that underwent the Raman transfer. $N$ is the BEC atom number. The polarization-qubit states $|L\rangle$ and $|R\rangle$ are mapped onto the magnon states $|2,\pm1\rangle$. This establishes the entangled atom-BEC state
\begin{equation}
\label{at-BEC}
|\psi_{\rm at\otimes BEC}\rangle = (|1,1\rangle\otimes|2,-1\rangle-|1,-1\rangle\otimes|2,1\rangle)/\sqrt2
.\end{equation}
All atoms in the BEC participate in the collective excitation, so that the single atom is entangled with an enormous number of atoms. The BEC serves as a quantum memory for one particle of an Einstein-Podolsky-Rosen pair \cite{clausen:11, saglamyurek:11}.

The entanglement is shared between two stand-alone experiments in different laboratories, with a physical separation of 13 m between the BEC and the single atom. Interfacing these two different atomic systems poses many experimental challenges. For example, optimizing the performance of each subsystem produces quite different values for the optimal light frequency, both on a coarse scale of $D_1$ line versus $D_2$ line and on a fine scale of tens of MHz. In addition, loading of a single atom must be synchronized with the production of a BEC and many write-read cycles must be performed on each BEC, to achieve sufficient average count rates. The large number of write-read cycles reduces the BEC atom number and along with it the average write-read efficiency.

In searching for the optimal photon wavelength, one challenge is off-resonant photoassociation driven by the EIT control laser because it gradually reduces the BEC atom number $N$. For $2\times 10^4$ write-read cycles per BEC, $N$ drops from $1.2\times 10^6$ to $0.2\times 10^6$ when operating near the atomic $D_1$ line at $\lambda=795$ nm. In principle, working at the $D_2$ line at $\lambda=780$ nm would be favorable, because here larger dipole matrix elements would result in larger optical depth of the BEC as well as higher efficiency $\epsilon$ of the single-photon generation in the cavity. In practice, however, the photoassociation rates are much higher at 780 nm, causing the atom number to decay 50 times faster. This forces us to work at 795 nm. For laser pulses at 795 nm, we still achieve a write-read efficiency of $\eta=(53\pm5)\%$ in the BEC, exceeding the 48\% reported previously \cite{novikova:07}.

The EIT control photons scattered off the BEC pose another problem, namely that a noticeable fraction of them reaches the photo detectors, producing such a large background that detecting the entanglement becomes impossible. We solve this problem by placing the photo detectors behind a single-mode fiber for transverse mode filtering and a filter cavity with a finesse of 180 for spectral filtering. The light retrieved from the BEC is so well collimated that we manage to couple $(58\pm3)\%$ of it into the single-mode fiber.

Yet another challenge arises from the fact that the optical dipole trap which holds the single atom inside the cavity shifts the atomic resonance by 130 MHz. Attempting to store photons at this shifted frequency would reduce the BEC write-read efficiency $\eta$. By tuning the trigger laser and the cavity, we produce photons with a blue detuning of 70 MHz relative to the free-space resonance. This choice is a tradeoff between reduced production efficiency $\epsilon$ and reduced BEC write-read efficiency $\eta$. The production efficiency $\epsilon$ is $(56\pm2)\%$ on resonance with the $D_2$ line, exceeding previous results in a similar system \cite{weber:09} by a factor of 6. Operation at 70 MHz detuning from the $D_1$ line reduces $\epsilon$ to $(14\pm1)\%$. Likewise, the BEC write-read efficiency is reduced to $\eta=18\%$ for laser pulses and to $\eta=(16\pm1)\%$ for single photons. The similarity of these two values for $\eta$ shows that possible noise on the carrier phase of the single-photon pulse is low enough to have negligible effect on $\eta$. Overall, we typically observe $\sim 2.5 \times 10^{-6}$ coincidences per experimental shot. For further experimental details see Ref.\ \cite{TheEPAPS}.

After establishing the matter-matter entanglement, we retrieve the single photon from the BEC. In addition, we generate a second single-photon pulse in the cavity, thereby mapping the atomic spin state onto the photon polarization. The generation of the second photon can be performed before, during, or after retrieving the first photon from the BEC. This produces the maximally-entangled two-photon singlet-state
\begin{equation}
\label{ph-ph}
|\psi_{\rm ph\otimes ph}\rangle
= |\psi^-\rangle
= (|R\rangle\otimes|L\rangle-|L\rangle\otimes|R\rangle)/\sqrt2
,\end{equation}
which is no longer entangled with the single atom or the BEC. Finally, the polarization of each photon is measured.

We perform many repetitions of a two-photon correlation measurement for 3 different settings of the polarization bases \cite{guehne:02} to determine the fidelity $F=\langle\psi^-|\rho|\psi^-\rangle$ of the experimentally produced density matrix $\rho$ with the Bell state $|\psi^-\rangle$. The Peres criterion $F>1/2$ is a sufficient condition for entanglement \cite{peres:96, horodecki:09}. We choose a storage time in the BEC of $t_{\rm BEC}= 1$ $\mu$s and a delay between the two single-photon pulses emitted from the cavity of $t_{\rm at}= 1$ $\mu$s. Our measurement yields an unbiased estimator and a statistical standard error of $F=(95.0\pm3.4)\%$. The likelihood function for $F$ is non-Gaussian. The confidence level for $F>1/2$ is $1-1.1\times 10^{-8}$, clearly showing the presence of entanglement. The observed value of $F$ is remarkably high, given that it characterizes the concatenation of four processes, namely creating the entanglement, mapping it to the BEC, mapping it back to a photon, and mapping the single-atom qubit onto the second photon. For comparison, we also performed the experiment without the BEC, so that only the atom-cavity system is characterized. This yields $F=(94.1\pm1.5)\%$ showing that a write-read cycle has no discernible effect on $F$.

\begin{figure}[t]
\includegraphics[width=0.9\columnwidth]{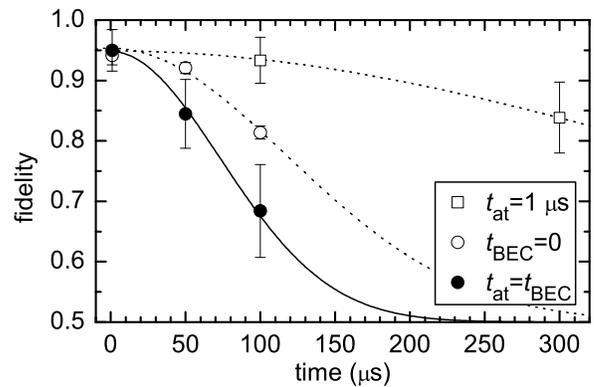}
\caption{
\label{fig-F}
{\bf Singlet-state fidelity.} The fidelity associated with the entangled state ($\bullet$) decays with a half-time of $86\pm18$ $\mu$s. The origin of this decay can be analyzed when considering additional data ($\circ$,$\square$) for which only one qubit is exposed to decoherence for a long time, whereas the other qubit is quickly measured. The single atom obviously decoheres faster than the BEC. The lines show Gaussian fits. The error bars are statistical standard errors.
}
\end{figure}

The matter-matter qubit entanglement in our system is found to exist for very long times. To observe a reduction of $F$ due to decoherence, one has to increase $t_{\rm at}$ and $t_{\rm BEC}$ drastically, as shown in Fig.\ \ref{fig-F}. In our setup, decoherence is dominated by local magnetic-field noise. To prevent this noise from redistributing population between Zeeman states, magnetic hold fields of 100 mG at the BEC and 40 mG at the single atom are applied. Assuming that the noise produces a Gaussian distribution of magnetic field values, we expect a Gaussian temporal decay of $F$ down to 1/2. The solid line in Fig.\ \ref{fig-F} shows a corresponding fit with a best-fit value for the entanglement half-time of $86\pm18$ $\mu$s. Note that the data point at $t_{\rm at}=t_{\rm BEC}=100$ $\mu$s still shows entanglement, with a confidence level of $97.1\%$ for $F>1/2$.

To guarantee that there is a time interval during which remote matter-matter entanglement exists, the first photon must be stored in the BEC before the second one is generated in the cavity. To this end, $t_{\rm at}$ and $t_{\rm BEC}$ must both exceed the sum $\tau\sim 0.6$ $\mu$s of the duration of a single-photon pulse and the time for photon transport. For $t_{\rm at}=t_{\rm BEC}=1$ $\mu$s, this is obviously the case. For $t_{\rm at}=t_{\rm BEC}=100$ $\mu$s, $t_{\rm BEC}/\tau$ is even as large as $\sim 200$.

To analyze the origin of the temporal decay of $F$, we also took data where either $t_{\rm at}$ or $t_{\rm BEC}$ was small and constant, so that only one of the atomic systems was exposed to decoherence for a variable time. Here, Gaussian fits yield half-times of $139\pm9$ $\mu$s for decoherence in the cavity ($\circ$) and $470\pm160$ $\mu$s for decoherence in the BEC ($\square$), corresponding to root-mean-square noise of 1.0 and 0.3 mG, respectively. While the atom-cavity system is the lifetime-limiting component in our present setup, the observed half-time of $F$ exceeds previous results from a similar system \cite{weber:09} by a factor of $\sim 20$.

\begin{figure}[t]
\includegraphics[width=0.95\columnwidth]{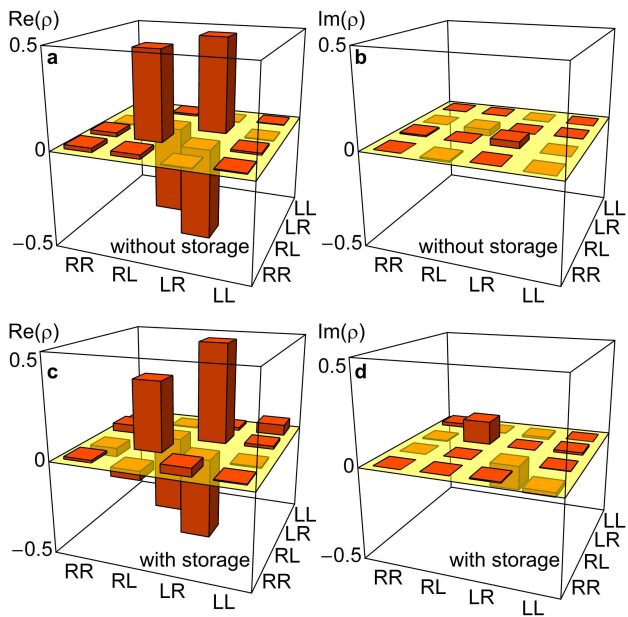}
\caption{
\label{fig-rho}
{\bf Reconstructed two-photon density matrix.} Real parts ({\bf a}, {\bf c}) and imaginary parts ({\bf b}, {\bf d}) of $\rho$ without ({\bf a}, {\bf b}) and with ({\bf c}, {\bf d}) storage in the BEC. Data were obtained with $t_{\rm at} = 1$ $\mu$s and with $t_{\rm BEC}=0$ or 1 $\mu$s. The density matrices are positive semidefinite and were obtained by a numerical maximum-likelihood search. Statistical errors for each matrix element are typically 0.01 without storage and 0.04 with storage, respectively.
}
\end{figure}

After showing that the system is entangled, we now turn to a reconstruction of the complete density matrix $\rho$ using quantum state tomography \cite{nielsen:00}. The experimental procedure is similar to the determination of $F$, but now the measurements need to be performed for 9 different settings of the polarization bases \cite{altepeter:05}. From such a measurement with $t_{\rm at}=t_{\rm BEC}=1$ $\mu$s, we reconstruct the density matrix shown in Fig.\ \ref{fig-rho} {\bf c},{\bf d}. The result closely resembles the expected singlet-state. For comparison we also performed a measurement to reconstruct $\rho$ without storage in the BEC, see Fig.\ \ref{fig-rho} {\bf a},{\bf b}. The comparison shows that the polarization state is maintained well during storage.

To conclude, we demonstrated remote entanglement of a single atom and a BEC in a scheme which has the potential to become deterministic. A high fidelity for a concatenation of four processes was achieved and the matter-matter entanglement was observed to exist for 100 $\mu$s. The efficiencies $\epsilon$ and $\eta$ used in the present experiment offer room for future improvements. On one hand, increasing the atom-cavity coupling, e.g.\ by reducing the cavity volume, would help to create single photons at the free-space resonance more efficiently. On the other hand, choosing a different alkali atom might result in a much lower scattering rate of EIT control photons, thus allowing for the use of the $D_2$ line with larger matrix elements and for more shots per BEC. Long-term opportunities lie in quantum networks \cite{kimble:08} and in quantum logic gates based on magnon-magnon interaction in the BEC \cite{rispe:1010.0037}. Furthermore, the atom-cavity system provides a clear perspective for the sequential generation of entangled multi-qubit states \cite{gheri:98, schon:05} and when combined with several BECs, it could be used to create a quantum network of several entangled BECs.

We thank D.\ Bauer and H.\ Specht for discussions and C.\ Guhl and A.\ Neuzner for assistance with the experiment. This work was supported by the German Excellence Initiative via NIM, by the European Union via the AQUTE program, by BMBF via IKT 2020, and by the DFG via SFB 631 and via Forschergruppe 635.

\section{Appendix}

\subsection{Time Evolution of Eq.\ \eqref{at-BEC}}

The energies of the two terms in Eq.\ \eqref{at-BEC} typically differ slightly in our experiment due to magnetic fields and mean-field energies. As a result, the relative phase of the two terms in Eq.\ \eqref{at-BEC} evolves in time.

A small magnetic hold field $B$ of less than 0.1 G along the quantization axis is deliberately applied in each setup to stabilize the spatial orientation of the atomic spins against magnetic field noise. The corresponding Zeeman energies produce a time dependence of the relative phase of the two terms in Eq.\ \eqref{at-BEC}. This relative phase is related to Larmor precession of the atomic spin or, equivalently, to Faraday rotation of the polarization of the light. As long as the magnetic field is temporally stable, the accumulated relative phase for a given storage time is fixed. We easily compensate it by an appropriate setting of the wave plates in front of each PBS. The decoherence observed in Fig.\ \ref{fig-F} is dominated by fluctuations of the magnetic fields in both setups.

Mean-field energies are created by two-atom collisions in the BEC. A single magnon with one atom in hyperfine state $|f,m_f\rangle$ and all other atoms in hyperfine state $|1,0\rangle$ experiences a mean-field energy that depends on the $s$-wave scattering length $a_{|1,0\rangle \otimes |f,m_f\rangle}$ for an elastic collision of one atom in state $|1,0\rangle$ with another atom in state $|f,m_f\rangle$. In the limit of vanishing magnetic field, the two scattering lengths $a_{|1,0\rangle \otimes |2,1\rangle}$ and $a_{|1,0\rangle \otimes |2,-1\rangle}$ for the magnons in hyperfine states $|2,1\rangle$ and $|2,-1\rangle$ become identical. This identity reflects symmetry under point reflection combined with the fact that no strong inelastic collision channels open or close at $B=0$ for these states \cite{kokkelmans:pers}. The magnetic hold field applied in our experiment is so small that the mean-field energy does not induce a noticeable shift of the relative phase of the two terms in Eq.\ \eqref{at-BEC} during the storage times explored here.

\subsection{Experimental Details}

The BEC has Thomas-Fermi radii of 7, 25, and 25 $\mu$m. The peak optical depth at the 780 nm cycling transition is 1500 at $N=1.2\times10^6$. The use of 795 nm light reduces the peak optical depth to 120. The EIT signal beam has a waist of 8 $\mu$m ($1/e^2$ radius of intensity) and propagates perpendicularly to the BEC symmetry axis. The EIT control laser is typically operated at a Rabi frequency of $\Omega_c/2\pi \sim 20$ MHz. At the peak optical depth, the width of the EIT transmission window in frequency space \cite{fleischhauer:05:appendix} is $\Delta\omega_\mathrm{trans}/2\pi\sim 6$ MHz.

After producing the BEC in state $|1,-1\rangle$, as in Ref.\ \cite{marte:02}, and transferring it to a dipole trap, we use two consecutive microwave $\pi$-pulses with different frequencies to transfer essentially all population first to $|2,0\rangle$ and then to $|1,0\rangle$. A blast light pulse removes remaining atoms with $f=2$ and transient application of a magnetic field gradient removes atoms in states $|1,\pm1\rangle$ from the shallow dipole trap. After this, we detect no atoms in undesired states, with a detection limit of $\sim 200$ atoms.

The cavity has a finesse of 56 000. For more details about the setup of the high-finesse cavity, see Ref.\ \cite{bochmann:09}.

\subsection{Count Rates}

An entangled photon pair is produced in the cavity with a probability of $(1.0\pm0.2)\% \approx\epsilon^2$. Successive attempts to load a single atom into the cavity yield an overall success probability of 72\% for a single atom to be present when the BEC is ready. Possible loss of the atom during $2\times10^4$ cycles reduces the average coincidence rate by a factor of 0.81. The BEC write-read efficiency is $\eta=16\%$. A 50:50 non-polarizing beam splitter distributes the photons between the laboratories, reducing the relevant coincidence rate by a factor of 0.25. All other beam transport (including stray light filtering) reduces the coincidence rate by a factor of 0.21. The avalanche photodiodes have a detection efficiency of typically 50\%. Combination of these factors should yield at best $1.2 \times 10^{-5}$ coincidences per shot. Routinely, we typically observe $\sim 2.5 \times 10^{-6}$ coincidences per shot. A new BEC is produced every 20 seconds. A typical data point in Fig.\ \ref{fig-F} is calculated from 60 coincidences, corresponding to $\sim 7$ hours of net data acquisition time per point.

\clearpage

\end{document}